\documentclass[%
 reprint,
 superscriptaddress,
 amsmath,amssymb,
 aps,
 prb,
]{revtex4-2}
\usepackage{graphicx}
\usepackage{dcolumn}
\usepackage{physics} 
\usepackage{bm}
\usepackage[dvipsnames]{xcolor}

\definecolor{darkblue}{rgb}{0, 0, 0.8} 
\definecolor{hred}{HTML}{E43E4C} %
\definecolor{ptpurple}{HTML}{AA3377}
\definecolor{ptgreen}{HTML}{228833}

\usepackage{silence}
\usepackage{hyperref}
\hypersetup{colorlinks=true, breaklinks=true, linkcolor=ptpurple, citecolor=ptpurple, urlcolor=ptpurple, pdftitle={Archimedean}} 

\def \beq{\begin{eqnarray}}
\def \eeq{\end{eqnarray}}

\def \r{{\bm r}}

\def \u{{\bm u}}

\def \k{{\bm k}}

\def \R{{\bm R}}

\def \ve{{\varepsilon}}

\newcommand{\nn}{\nonumber \\}

\usepackage{lipsum} 

\newcommand{\mb}{\mathbf} 

\begin{document}

\title{ Ground states of quantum XY dipoles on the Archimedean lattices}

\newcommand{\HarvardPhysicsAddress}{Department of Physics, Harvard University, Cambridge, Massachusetts 02138, USA}
\newcommand{\BerkeleyPhysicsAddress}{Department of Physics, University of California, Berkeley, CA 94720, USA}
\newcommand{\LBNLAddress}{Material Science Division, Lawrence Berkeley National Laboratory, Berkeley, CA 94720, USA}
\newcommand{\TUMAddress}{Technical University of Munich, TUM School of Natural Sciences,
Physics Department, James-Franck-Str. 1, 85748 Garching, Germany}
\newcommand{\MunichAddress}{Munich Center for Quantum Science and Technology (MCQST), Schellingstr. 4, 80799 M\"unchen, Germany}
\newcommand{\CMUAddress}{Department of Physics, Carnegie Mellon University, Pittsburgh, PA 15213, USA}

\author{Marcus Bintz$^*$}
\affiliation{\HarvardPhysicsAddress}

\author{Ahmed Khalifa$^*$}
\affiliation{\CMUAddress}

\author{Vincent S. Liu}
\affiliation{\HarvardPhysicsAddress}

\author{Johannes Hauschild}
\affiliation{\TUMAddress}
\affiliation{\MunichAddress}

\author{Michael P. Zaletel}
\affiliation{\BerkeleyPhysicsAddress}
\affiliation{\LBNLAddress}

\author{Shubhayu Chatterjee}
\affiliation{\CMUAddress}

\author{Norman Y. Yao}
\affiliation{\HarvardPhysicsAddress}

\date{\today}
             
\begin{abstract}    
    We report numerical ground states for the dipolar XY spin model, which describes extended antiferromagnetic interactions in two-dimensional arrays of polar molecules and two-level Rydberg atoms.   Carrying out large-scale density matrix renormalization group (DMRG) calculations, we compute ground state properties on nine of the eleven Archimedean lattices---tilings of the plane by regular polygons.   Four of these host trivial paramagnets, while another four develop collinear Ne\'el magnetic order, as was found previously for the square lattice. For the ordered states, we calculate the hydrodynamic parameters (magnetization, susceptibility, and stiffness) and compare to linear spin wave theory. We also investigate the triangular lattice, for which we find several competing phases including coplanar magnetism, stripe density wave order, and a possible spin liquid; their relative stability is sensitive to the long-range couplings present in our dipolar model. Finally, the Archimedean classification is completed by the kagome lattice, which we argue in a companion work is likely to be a Dirac spin liquid. 
\end{abstract}
\maketitle

\section{Introduction}

Geometry is essential to matter: two collections of microscopic particles with the same elemental composition and fundamental interactions, differing only in their spatial arrangement, will often have completely different macroscopic material properties.
Even minimalistic systems exhibit this sensitivity; a well-known example is the Heisenberg model of $S=1/2$ spins coupled by local spin-exchange on a two-dimensional lattice~\cite{richterQuantumMagnetismTwo2004}.
When placed on the vertices of square tiling, such spins develop conventional, collinear antiferromagnetic order at low temperature~\cite{manousakisSpin1/2HeisenbergAntiferromagnet1991,whiteNeelOrderSquare2007}.
But if one simply begins with a different tiling of the plane---such as with hexagons and corner-sharing triangles---wholly distinct phases of matter can emerge, possibly including exotic quantum spin liquids~\cite{elserNuclearAntiferromagnetismRegistered1989, misguichTWODIMENSIONALQUANTUMANTIFERROMAGNETS2013, 
yanSpinLiquidGroundState2011, savaryQuantumSpinLiquids2017}.

Geometry is also increasingly a domain under experimental control.
One especially exciting area of development is in ultracold atom and molecule experiments, where it is now possible to orchestrate large arrays of individually-positioned, tightly focused laser beams that each trap single particles
~\cite{kimSituSingleatomArray2016,barredoAtombyatomAssemblerDefectfree2016, endresAtombyatomAssemblyDefectfree2016, kaufmanQuantumScienceOptical2021}.
In such highly-configurable optical tweezer arrays~\cite{linAIEnabledParallelAssembly2025, chiuContinuousOperationCoherent2025, holmanTrappingSingleAtoms2026}, the lattice spacing is typically on the order of several microns---this has key consequences for the natural interactions at play.
Namely, tunneling-induced exchange is exponentially suppressed, and the most relevant inter-particle couplings are instead mediated by the ambient electromagnetic field~\cite{browaeysManybodyPhysicsIndividually2020}.
These can be maximized by employing particles with large local moments, e.g. polar molecules or Rydberg atoms.

A native Hamiltonian commonly arising in such experiments is the spin-1/2 dipolar XY model,
\begin{equation}\label{eq:HdXY}
    H_{\rm{dXY}} =  J \sum_{i<j} \frac{S_i^+S_j^- + S_i^-S_j^+}{(r_{ij}/a)^3} 
\end{equation}
which describes the dipole-dipole interaction~\cite{hazzardQuantumCorrelationsEntanglement2014} between atomic~\cite{lippeExperimentalRealization3D2021, bornetScalableSpinSqueezing2023a,geierTimereversalDipolarQuantum2024} or molecular~\cite{baoDipolarSpinexchangeEntanglement2023, liTunableItinerantSpin2023,ruttleyLonglivedEntanglementMolecules2025, hollandOndemandEntanglementMolecules2023, christakisProbingSiteresolvedCorrelations2023c} qubits possessing a large internal transition dipole moment, but no permanent moment (i.e. due to parity symmetry).
Here, $r_{ij}$ is the distance between spins $i$ and $j$, $a$ is a microscopic length scale (lattice spacing), and $J$ is the coupling constant; we focus here on the antiferromagnetic case, $J>0$.
The dipolar XY Hamiltonian has no free parameters, and so its ground state is completely determined by the geometric arrangement of the atoms or molecules.
We are thus motivated to investigate the geometry-dependence of the $H_{\rm{dXY}}$ ground state.
Experimentally, there is already evidence that a simple square lattice arrangement leads to an ordered state, spontaneously breaking the continuous $U(1)$ symmetry generated by the conserved magnetization $M_z = \sum_i S_i^z$~\cite{chenContinuousSymmetryBreaking2023,peterAnomalousBehaviorSpin2012, sbierskiMagnetismTwodimensionalDipolar2024}.
By contrast, in a companion work we numerically show that on the kagome lattice, strong geometric frustration leads to a disordered state, likely to be the gapless Dirac spin liquid~\cite{bintzDiracSpinLiquid2024}.
Both of these two-dimensional geometries are examples of so-called Archimedean lattices, formed by tilings of regular polygons where all vertices are symmetry-equivalent~\cite{richterQuantumMagnetismTwo2004, grunbaumTilingsPatterns2016}.
There are eleven such planar arrangements---in this paper, we study the ground state of $H_{\rm{dXY}}$ on the remaining nine.

Carrying out large-scale density matrix renormalization group (DMRG) calculations~\cite{whiteDensityMatrixFormulation1992,stoudenmireStudyingTwoDimensionalSystems2012,hauschildEfficientNumericalSimulations2018}, we find these states fall into two general classes, with one exception. 
The first common phase is collinear $U(1)$ symmetry breaking order,  evinced by long-range $\langle S^x_i S^x_j \rangle$ correlations with a clear Ne\'el sign structure.
This occurs when frustration is weak: on the square, hexagonal, truncated square, snub square, and truncated trihexagonal lattices (Sec.~\ref{sec:AFM}).
In these cases, the order can be captured analytically by semiclassical linear spin wave theory; this has three characteristic hydrodynamic parameters, which we can also compute with DMRG (Sec.~\ref{sec:hydro}).
A second possibility arises for more frustrated systems but with an even number of spins per unit cell: the elongated triangular, truncated hexagonal,  rhombitrihexagonal,  and snub trihexagonal lattices  (Sec.~\ref{sec:singlet}).
On these geometries, neighboring spins form local two- or six-spin singlet states, and the full many-body wavefunction is a trivial paramagnet, i.e.~to good approximation, a tensor product of the local singlets.
Finally, for the triangular lattice we find that several phases closely compete in energy, including conventional coplanar order and a possible spin liquid (Sec.~\ref{sec:tri}).

\begin{figure}
    \includegraphics{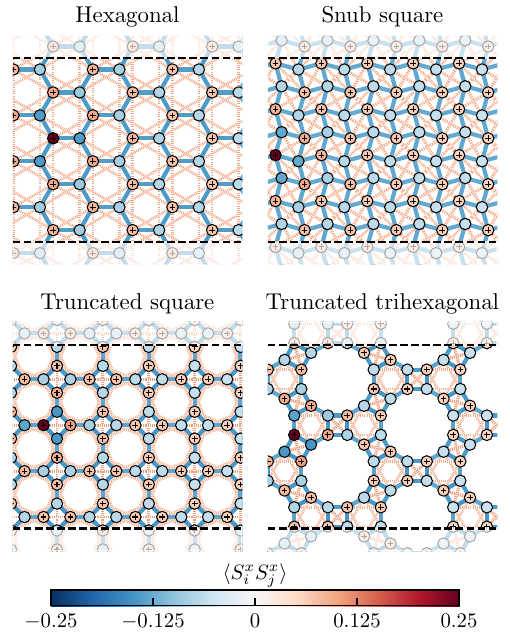}
    \caption{\label{fig:collinear} Depiction of the collinear-ordered ground states of $H_{\rm{dXY}}$ on four of the Archimedean lattices, as found in iDMRG ($d=2048$).  Bonds show correlations between nearest-neighbor and next-nearest neighbors spins; red dashed lines correspond to ferromagnetic $\langle S^x_i S^x_j\rangle>0$, while blue solid lines are antiferromagnetic.  Circles show $\langle S^x_i S^x_0\rangle$ correlations with a fixed site (dark red circle), and those with $\langle S^x_i S^x_0 \rangle >0$ are also marked with a '$+$' sign.  
    }
\end{figure}

\section{Collinear antiferromagnets}\label{sec:AFM}

In this section, we introduce the four lattices which numerically are found to have collinear antiferromagnetic order.
We first present the results from infinite DMRG (iDMRG) calculations, and then discuss our formalism (linear spin wave theory) for analytically understanding this order.
The quantitative calculation of the  characteristic hydrodynamic parameters for these orders is done afterwards in Section~\ref{sec:hydro}.

\subsection{iDMRG}\label{sec:AFM_iDMRG}
%
%
We compute the ground state of $H_{\rm{dXY}}$ with iDMRG, making use of the TeNPy library~\cite{hauschildEfficientNumericalSimulations2018}.
In order to apply the one-dimensional iDMRG algorithm, we compactify the plane into a cylinder with infinite length, and finite circumference $W$.
Additionally, the full long-range interactions cannot be exactly represented in the iDMRG setting; there is also some ambiguity in how the multiple paths around the cylinder should be accounted for.
We choose to simply truncate, so that only sites with distance $r_{ij} <  R_{\mathrm{max}} = W/2$ are coupled, and the coupling strength only includes the contribution from the shortest-distance path.
This and other choices of prescription for the interactions converge in the thermodynamic $W\to\infty$ limit, but at finite $W$ there can be quantitative and possibly qualitative differences in the ground state properties.
One generic possibility for the ground state is symmetry-breaking order: $H_{\rm dXY}$ is invariant under $U(1)$  spin rotations generated by the $z$-magnetization, $M^z = \sum_i S_i^z$,  rotations by $\pi$ about the $x$ or $y$ spin axes, time reversal, and any spatial symmetries of the lattice.
Here, we consider states in the $M^z = 0$ sector, employing spin-conserving tensors that respect the $U(1)$ symmetry.
The basic signature of spontaneous $U(1)$ symmetry breaking in this setting is off-diagonal long-range order in the $\langle S_i^x S_j^x \rangle = \langle S_i^y S_j^y \rangle $ correlations~\cite{yangConceptOffDiagonalLongRange1962}.
Five of the lattices we checked exhibited this type of order: square (not shown), hexagonal (``honeycomb''), snub square, truncated square, and truncated trihexagonal (Fig.~\ref{fig:collinear}c). 
In these lattices, the ground state $\langle S_i^x S_j^x \rangle$ correlations are large and roughly constant out to long distances, before slowly decaying due to the quasi-1d infinite cylinder geometry and the finite MPS bond dimension.
The correlations exhibit an obvious sign structure characteristic of a simple, bipartite Ne\'el antiferromagnet. 
Such order is frustrated at the nearest-neighbor level for the snub square tiling, but only at longer distances (and hence weaker coupling) in the other four cases.
%
%

\subsection{Linear spin wave theory}\label{sec:AFM_LSWT}

For the bipartite N\'eel antiferromagnets, we expect that a semi-classical description using linear spin wave theory (LSWT) to be reasonably accurate~\cite{andersonApproximateQuantumTheory1952, auerbachInteractingElectronsQuantum1994}.
Therefore, we briefly discuss the formalism for LSWT, and detail how to compute the long-wavelength hydrodynamic parameters~\cite{halperinHydrodynamicTheorySpin1969, hasenfratzFiniteSizeTemperature1993,sandvikGroundstateParametersFinitesize1999}, namely, the magnetization $m$, the out-of-plane susceptibility $\chi_\perp$ and the spin-wave velocity $v$ of the acoustic mode which in turn leads to the spin-stiffness $\rho_s = v^2 \chi_\perp$, enabling a comparison with the DMRG results. 

The starting point of LSWT is the fully polarized N\'eel antiferromagnet, i.e., $\langle S^{x} \rangle = \sigma S$, where $\sigma = \pm 1$ on sublattices of type $A$ and type $B$ respectively.
We denote the number of each type in the unit cell as $s$ (i.e. each unit cell has $2s$ sites).
We will soon supplement this, in our computation of $\chi_\perp$, by applying an out-of-plane magnetic field $h$,
\begin{equation}
 H = H_{\rm dXY} - h \sum_i S^z_i ,
\end{equation}
which leads to a uniform canting of all the spins along $\hat{z}$, i.e. a nonzero expectation value $m_z = \langle S^{z}_i \rangle$.
Hence, it is convenient to develop LSWT in the presence of this field and set $h = 0$ to find the other hydrodynamic quantities like $m$ or $\rho_s$.

To organize our analytic calculations, we first set some detailed notation for the lattice.
We index each site with dual indices $i\ \equiv (n,\eta)$, where $\R_n$ is Bravais lattice site index and $\u_\eta$ with $\eta \in \{1, 2, \cdots, 2s \}$ being the sublattice index, such that $\r_i = \r_{(n,\eta)} = \R_n + \u_\eta$. 
In this notation, we may re-write the Hamiltonian as
\begin{equation}
H = J \sum^\prime_{(n,\eta),(n^\prime,\eta^\prime)} \frac{S^{x_0}_{n,\eta} \, S^{x_0}_{n^\prime,\eta^\prime} + S^{y_0}_{n,\eta} \, S^{y_0}_{n^\prime,\eta^\prime}}{|\R_n + \u_\eta - \R_{n^\prime} - \u_{\eta^\prime}|^3} - h \sum_{n,\eta} S^{z_0}_{n,\eta}
\label{eq:Hh0}
\end{equation}
where we have introduced the lab-fixed frame $(x_0, y_0, z_0)$ for future convenience, and $\sum^\prime$ indicates that we avoid the same site $(n, \eta) = (n^\prime,\eta^\prime)$ in the summation.
It is also convenient to generalize the Fourier transform of the dipolar sum to a matrix in sublattice space:
\begin{equation}
\ve_{\k, \eta \eta^\prime} = -\sum^\prime_n \frac{\sigma_\eta \sigma_{\eta^\prime}}{|\R_n + \u_\eta  - \u_{\eta^\prime}|^3}  e^{- i \k \cdot \R_n}
\end{equation}
where the summation runs over all the underlying Bravais lattice points $\R_n$ except the origin if $\eta$ and $\eta^\prime$ correpond to the site within the unit cell, and all points $\R_n$ otherwise. 
As before, $\sigma_\eta$ indicates the staggered sign, i.e., $\sigma=1$ for $\eta\in A$ and $\sigma=-1$ for $\eta\in B$.
We note that $\ve_{\k, \eta \eta^\prime}$ is periodic in the Brillouin Zone (BZ) by definition, and satisfies $\ve_{\k, \eta \eta^\prime} = \ve_{-\k, \eta^\prime \eta} = \ve^*_{\k, \eta^\prime \eta}$.
Finally, we define $\ve_0 = \sum_{\eta^\prime} \ve_{\k= 0,\eta,\eta^\prime}$, which characterizes the dipolar interaction energy of one spin with all the other spins in the system in the in-plane polarized N\'eel state, and is accordingly indepedent of the sublattice index $\eta$.

In the presence of $h \neq 0$, it is energetically favorable for the spins to cant uniformly along the $z_0$ direction~\cite{auerbachInteractingElectronsQuantum1994}.
Taking this canting into account, we move to a twisted reference frame where we locally rotate the axes about $\hat{y}_0$ on each sublattice such that the N\'eel vector in the canted phase points along $\hat{x}$ in the new frame (i.e., the spins point along $\hat{x}$ in sublattice A and along $-\hat{x}$ in sublattice B).
The spins transform as,
\begin{equation}
\begin{pmatrix}
S^x \\
S^y \\
S^z
\end{pmatrix}_{n,\eta} = \begin{pmatrix}
\cos \theta_\eta & 0 & \sin \theta_\eta \\
0 & 1 & 0 \\
- \sin \theta_\eta & 0 & \cos \theta_\eta
\end{pmatrix} \begin{pmatrix}
S^{x_0} \\
S^{y_0} \\
S^{z_0}
\end{pmatrix}_{n,\eta}, 
\vspace{0.5pt}
\end{equation}
with $\theta_\eta = \sigma_\eta  \theta$ captures the staggering of in-plane components of the spins.

We fix the angle $\theta$ by demanding that the effective field along the (rotated) $z$ direction vanishes in the Hamiltonian re-written in the twisted reference frame, finding (see SM~\cite{supp} for a derivation)
\begin{equation}
\sin \theta = \frac{h}{2 J S \ve_0}.
\label{eq:sth}
\end{equation}
Now, we define the linearized Holstein-Primakoff (HP) bosons $a_{n,\eta}$ (in the twisted reference frame) as follows:
\begin{align}
S^x_{n,\eta} &= \sigma_\eta(S - a^\dagger_{n,\eta} a_{n,\eta}), \nonumber \\
S^y_{n,\eta} &= \frac{\sqrt{2S}}{2i}(a^\dagger_{n,\eta} - a_{n,\eta})\sigma_\eta, \nonumber \\
S^z_{n,\eta} &=  \frac{\sqrt{2S}}{2}(a^\dagger_{n,\eta} + a_{n,\eta}),
\label{eq:HPdef}
\end{align}
such that their Fourier transforms are given by:
\begin{equation}
a_{n,\eta} = \frac{1}{\sqrt{N}} \sum_\k e^{i \k \cdot \R_n} a_{\k,\eta}
\end{equation}
where $N$ is the total number of unit cells, and the sum on $\k$ runs over the BZ of the underlying Bravais lattice.
We are finally in a position to write the Hamiltonian in Eq.~\eqref{eq:Hh0} in a quadratic form which can be readily diagonalized.
\begin{widetext}
\begin{align}
H &= H^{(0)}_{\rm cAFM} + H^{(2)}, \text{ where }
H^{(0)}_{\rm cAFM} = -2sN \left[ J S(S + 1) \ve_0 + \frac{h^2}{4 J \ve_0} \right] \nn
H^{(2)} & = \frac{J S}{2} \sum_{\k,\eta, \eta^\prime} \left[ 2 \ve_0 \, \delta_{\eta \eta^\prime} - (1 + \sin^2\theta) \ve_{\k,\eta, \eta^\prime} \right] \left(  a^\dagger_{\k,\eta} a_{\k,\eta^\prime}  +   a_{-\k,\eta} a^\dagger_{-\k,\eta^\prime} \right) + \cos^2\theta \, \ve_{\k, \eta \eta^\prime} 
 \left( a_{-\k,\eta} a_{\k, \eta^\prime} +  a^\dagger_{\k, \eta} a^\dagger_{-\k,\eta^\prime}  \right)
 \label{eq:HHPquadratic}
\end{align}
\end{widetext}
We note that there is an O$(h^2/\ve_0)$ reduction of the energy per site in $H^{(0)}_{\rm cAFM}$, corresponding to the energy gained by canting to partially align with the applied field.  

To find the excitation spectrum and other physical quantities of interest, our task is to diagonalize $H^{(2)}$, for which we move to the Nambu (particle-hole) basis:
\begin{align}
H^{(2)} &= \frac{1}{2} \sum_\k \Psi^\dagger_\k h_\k \Psi_\k, \text{ where } \nonumber \\
\Psi_\k &= \left(
    a_{\k,\eta_1},
    a_{\k, \eta_2},
    \cdots 
    a_{\k, 2s},
    a^\dagger_{-\k,\eta_1},
    a^\dagger_{-\k,\eta_2},
    \cdots,
    a^\dagger_{-\k,2s}
\right)^T
\end{align}
is a Nambu spinor of dimension $4s$, and $h_\k$ is a $4s \times 4s$ matrix that we need to diagonalize. 
To do so while preserving canonical commutation relations,  we consider a para-unitary transformation~\cite{colpaDiagonalizationQuadraticBoson1978,xiaoTheoryTransformationDiagonalization2009} $\Psi_\k = W_\k \Gamma_\k$, such that 
\begin{align}
\Gamma_\k = \left(
    \gamma_{\k, 1}, \gamma_{\k, 2}, \cdots, \gamma_{\k, {2s}}, \gamma^\dagger_{-\k,1}, \gamma^\dagger_{-\k, 2},
    \cdots, \gamma^\dagger_{-\k, {2s} }
\right)^T.
\end{align}
The matrix $W_\k$ is obtained by diagonalizing the so-called dynamical matrix $D_\k = \tau^3 h_\k$ which defines the Heisenberg equations of motion for $\Psi_\k$. 
Here, $\tau^3$ is the diagonal Pauli-$Z$ matrix in Nambu space. 
The matrix $W_\k$, which satisfies $W_\k \tau^3 W_\k^\dagger = \tau^3$,  simultaneously diagonalizes both $h_\k$ and $D_\k$, such that 
\begin{equation}
W_\k^\dagger h_\k W_\k = \Omega_\k = \text{diag}\{ \omega_{\k, 1}, \cdots, \omega_{\k,2s},\omega_{\k, 1}, \cdots, \omega_{\k,2s} \},
\end{equation}
and
\begin{align}
 W_\k^{-1} D_\k W_\k &= \tau^3 \Omega_\k  \nonumber \\
 &= \text{diag}\{ \omega_{\k, 1}, \cdots, \omega_{\k,2s},-\omega_{\k, 1}, \cdots,-\omega_{\k,2s} \},
\end{align}
where $\mu \in \{1,2,\cdots,2s\}$ is the band index for the spin-wave excitations,  and $\omega_{\k,\mu} \geq 0$ is the spin-wave dispersion in the $\mu^{th}$ band.
Accordingly, the quadratic part of the Hamiltonian can be written in its diagonal basis as
\begin{equation}
H = H^{(0)}_{\rm cAFM} + \frac{1}{2} \sum_{\mu = 1}^{2s} \sum_\k \omega_{\k,\mu} + \sum_ {\mu=1}^{2s} \sum_{\k} \omega_{\k,\mu} \gamma^\dagger_{\k, \mu} \gamma_{\k, \mu}.
\end{equation}

Next, we discuss how to compute the different hydrodynamic observables, focusing on the $T = 0$ limit for a direct comparison with iDMRG.
First, to find the reduction of the ordered moment from its classical value of $S$ due to quantum fluctuations, we simply set $h = 0$ such that the twisted frame and the lab-fixed frame coincide, and calculate
\begin{equation}
m = \sigma_\eta \langle S^{x_0}_{n,\eta} \rangle = \sigma_\eta \langle S^{x}_{n,\eta} \rangle = S - \langle a^\dagger_{n,\eta} a_{n,\eta} \rangle
\end{equation}

Second, to compute the susceptibility $\chi_\perp$, we consider $h \neq 0$, and calculate the total magnetization $M_{z_0}$ in the $z_0$-direction induced by $h$:
\begin{align}
M_{z_0} &= \sum_{i} \langle S^{z_0}_i \rangle = \sum_{(m,\eta)} \left( \langle S^z_{m,\eta} \rangle\cos\theta_\eta  + \langle S^x_{m,\eta} \rangle \sin\theta_\eta \right) \nonumber \\ 
& = \frac{h}{2 J S \ve_0} \sum_{(m,\eta)} (S - \langle a^\dagger_{m,\eta} a_{m,\eta}\rangle ) 
\label{eq:chiU}
\end{align}
where we have used that $\langle S^z_{m,\eta} \rangle = 0$ within LSWT as it is made of operators that are linear in HP bosons, and that $\sin \theta_\eta = \sigma_\eta  \sin\theta  = \sigma_\eta h/(2 J S \ve_0)$. 
The first term in Eq.~\eqref{eq:chiU} is just the classical out-of-plane magnetization due to canting induced by $h$, and the second term represents its reduction due to quantum fluctuations. 
Since we only want the susceptibility as $h \to 0$, we can evaluate $ \langle a^\dagger_{m,\eta} a_{m,\eta}\rangle$ for $h = 0$ since the overall proportionality constant  already scales as $h$.  
In the SM~\cite{supp}, we show that the out-of-plane susceptibility susecptibility (per spin) from Eq.~\eqref{eq:chiU} can be written in terms of the para-unitary matrix $W_\k$ which diagonalizes $H^{(2)}$ as
\begin{align}
\chi_\perp &= \frac{1}{2sN} \frac{\partial M_{z_0}}{\partial h}\bigg|_{h \to 0} \nonumber \\
&= \frac{1}{2 J S \ve_0}\left(S + \frac{1}{2} - \frac{1}{4sN} \sum_{\k} \sum_{\nu = 2s+1}^{4s} [W_\k^\dagger W_\k]_{\nu,\nu}   \right)
\label{eq:chip}
\end{align}
where $2sN$ is the total number of sites. The leading term, $\chi_\perp \approx  1/(2 J \ve_0)$ gives the classical susceptibility, the rest are quantum corrections from LSWT which reduce the size of the ordered moment and hence suppresses the susceptibility. 

Finally, to find the spin-wave velocity, we again set $h = 0$ and focus on the acoustic spin-wave branch $\mu = \mu_0$ which is gapless at the $\Gamma$-point. 
Generally, $\omega_{\k, \mu = \mu_0} \approx v |\k|$ for small $|\k|$, with $v$ being the corresponding spin-wave velocity which is numerically extracted by fitting the dispersion at low momenta. 

To conclude this section, we note that the spin-wave dispersion obtained from LSWT (at $h=0$) has a minimum at the $\Gamma$-point in the BZ for the square, hexagonal, truncated square and truncated trihexagonal lattices, as expected for a N\'eel antiferromagnet.
However, for the snub square lattice, we find that the spin-wave energy goes to zero at a finite wave-vector (see SM for a plot of the spin-wave dispersion assuming a N\'eel phase) and becomes imaginary near the $\Gamma$ point, indicating an instability of the N\'eel state~\cite{lifshitzStatisticalPhysicsTheory2013}.
To investigate this further, we carry out a Luttinger-Tisza analysis~\cite{luttingerTheoryDipoleInteraction1946} of the ordering wave-vector~\cite{supp}.
Within this framework, the classical ground state is found by diagonalizing the Fourier transform $\tilde{J}(\k)$ of the real space interaction matrix $J_{ij}$ between spins: the wave-vector $\k_{\rm LT}$ corresponding to the minima determines the ordering pattern. 
We indeed find the N\'eel state to be the classical ground state for the four lattices where the LSWT results are meaningful. 
%

\begin{figure}
    \includegraphics{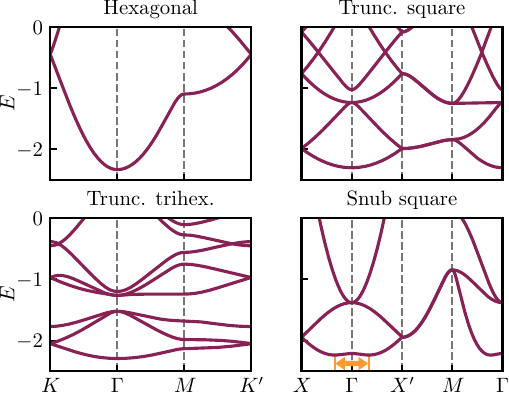}
    \caption{\label{fig:LT_spectrum}
    Luttinger-Tisza spectra of the four ordered lattices. The classical energy of the hexagonal, truncated square, and truncated trihexagonal lattices are minimized at the $\Gamma$ point. However, for the snub square this minimum is split and displaced (highlighted by orange arrows), suggesting a non-N\'eel state at the classical mean-field level.
    }
\end{figure}

%
On the other hand, as shown in Fig.~\ref{fig:LT_spectrum}, we find that the minimum of the Luttinger-Tisza spectrum occurs slightly displaced from the $\Gamma$ point for the snub square lattice, at some small symmetry-related incommensurate wave-vectors. 
This hints at the possibility that the snub square lattice can have a ground state that is different from the N\'eel state, which renders LSWT with the N\'eel state as the starting point infeasible.  
As for our numerical observations: at the DMRG-accessible system sizes, we always find a simple N\'eel-like order for the snub square ground state, but we believe it is rather plausible that the incommensuration instability only appears at larger length scales.
Alternatively, it may be that strong quantum effects ultimately favor the N\'eel order over a classical spiral~\cite{chernyshevQuantumStabilizationUnexpected2025}.

\section{Hydrodynamic parameters}\label{sec:hydro}

As discussed above, the low-energy physics of the collinear $U(1)$ symmetry-breaking order phases can be described by a single-component spin wave theory.
This hydrodynamic theory carries three free parameters, which, in a DMRG settting, are conveniently chosen as the magnetization, $m$, spin susceptibility, $\chi_\perp$, and spin stiffness, $\rho$.
Other characteristic quantities of this phase follow; in particular, the spin-wave velocity is given by $v = \sqrt{\rho/\chi_\perp}$~\cite{halperinHydrodynamicTheorySpin1969}.
In this section we compute these values for the $H_{\rm{dXY}}$ ground state on the four ordered lattices.
Compared to short-range models, the $1/r^3$ interactions (i) restrict the the range of system sizes which are computationally accessible, and (ii) enhance finite-size effects on the systems where we are able to reach good convergence.
Consequently, the precision with which we can estimate the thermodynamic limit of these values is generally unclear.
As a comparison point for the DMRG results, we will also report values analytically computed in linear spin wave theory (LSWT).
%

\subsection{Magnetization}\label{sec:hydro_mag}

\begin{figure}
    \includegraphics{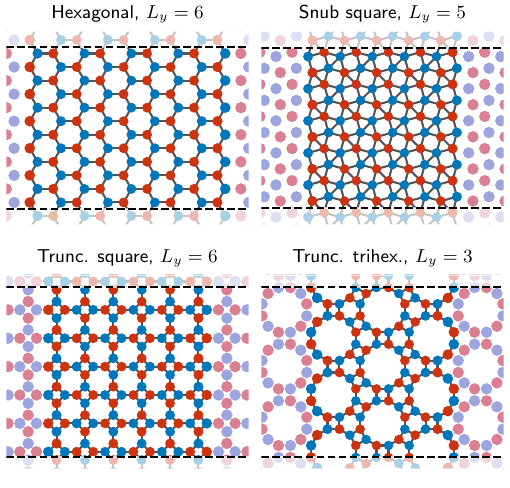}
    \caption{\label{fig:maggeometry}
    Finite cylinder segments in the presence of a $U(1)$-symmetry breaking environment.  Circles show $\langle S^x \rangle$ for each spin, with area proportional to $|\langle S^x \rangle |$ and color indicating the sign (red for positive, blue for negative).  Environment spins (pale circles) are fully polarized, $\langle S^x \rangle = \pm 1/2$. The center-segment spins show the local magnetization of the DMRG ground state, subject to the field from the external spins.
    }
\end{figure}

\begin{figure}
    \includegraphics{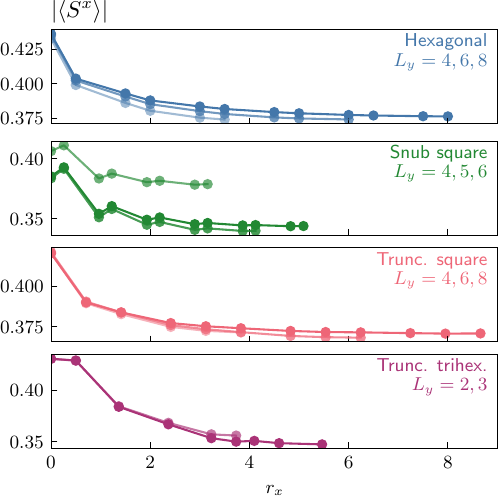}
    \caption{\label{fig:magprofile}
    Magnetization profile $\langle S^x (r_x)\rangle $ of the symmetry-broken ground states, with $r_x$ the linear distance from the leftmost spin of the finite segment. We average over spins with the same $r_x$, and plot only the left half of the profile; the right half is mirror-symmetric, up to differences of order the MPS truncation error, $\epsilon_t$ (see Table~\ref{tab:magnetization}). Different weight lines show the profile for varying cylinder sizes, with fixed aspect ratio $L_y/L_x$.  The magnetization of the central spins (i.e. largest shown $r_x$) is recorded in Table~\ref{tab:magnetization}.
    }
\end{figure}

\begin{table}
\caption{\label{tab:magnetization}Magnetization of finite cylinder segments.}
\begin{ruledtabular}
\begin{tabular}{llld|d|ll}    
Lattice & $L_x$ & $L_y$ & W & m & $d$ & $\epsilon_t$\\
\hline
Hexagonal & 6 & 4 & 6.93 & 0.374 & 1024 & $2 \times 10^{-8}$ \\
  & 9 & 6 & 10.39 & 0.374 & 1024 & $4 \times 10^{-7}$ \\
  & 12 & 8 & 13.86 & 0.376 & 1024 & $4 \times 10^{-6}$ \\
   & & & \multicolumn{1}{c|}{LSWT} & 0.384 & &  \\
  \hline
Snub square & 4 & 4 & 7.73 & 0.378 & 1024 & $6 \times 10^{-7}$ \\
  & 5 & 5 & 9.66 & 0.340(3) & 1024 & $9 \times 10^{-6}$ \\
  & 6 & 6 & 11.59 & 0.344(2) & 1536 & $1 \times 10^{-5}$ \\
  \hline
Trunc. square & 4 & 4 & 9.66 & 0.372 & 1024 & $4 \times 10^{-8}$ \\
  & 6 & 6 & 14.49 & 0.368 & 1024 & $3 \times 10^{-6}$ \\
  & 8 & 8 & 19.31 & 0.371(1) & 1024 & $2 \times 10^{-5}$ \\
     & & &  \multicolumn{1}{c|}{LSWT} & 0.373 & &  \\

  \hline
Trunc. trihex. & 2 & 2 & 9.46 & 0.368 & 1024 & $3 \times 10^{-7}$ \\
  & 3 & 3 & 14.20 & 0.347 & 2048 & $3 \times 10^{-5}$ \\
       & & &  \multicolumn{1}{c|}{LSWT} & 0.364 & &  \\

\end{tabular}
\end{ruledtabular}
\end{table}

We first determine the bulk magnetization $m$ by computing the ground states of finite-size systems in the presence of a symmetry-breaking environment~\cite{whiteNeelOrderSquare2007}.
As depicted in Fig.~\ref{fig:maggeometry}, we take our system to be a finite-length $L$ central segment of a long cylinder of circumference $W$.
The environment ($E$) spins outside this segment are fully pinned, $\langle S^x_E\rangle = \pm 1/2$, and act as a source of a non-uniform symmetry-breaking field for the system through the dipolar XY interaction.
In practice, we represent only the system spins, and use finite DMRG to find their ground state under $H_{\rm{dXY}} + \sum_i h_i S^x_i$, where $h_i = 2\sum_{j\in E} \langle S^x_j\rangle / r_{ij}^3$~\footnote{In this sum, we only include environment spins with $r_{ij}<R_{\mathrm{max}}$, using the same cutoff radius $R_{\mathrm{max}} = W/2$ as for the intrasystem spin-spin interactions.}.
With this setup, $m$ is estimated as simply the expectation value $\langle S_i^x \rangle$ of a central system spin away from the environment (averaged over the ring at the center of the segment). 
In Fig.~\ref{fig:magprofile}, we show the DMRG spatial profile of the local magnetization as a function of the linear distance from the left edge; the magnetization at the edge is generically larger than in the bulk, simply due to the environment-generated magnetic field.
In the thermodynamic limit $W\to \infty$ (with aspect ratio $L/W$ fixed), the bulk $\langle S^x_i\rangle$ and other definitions of $m$ should converge to the same value.

The results of our calculations are summarized in Table~\ref{tab:magnetization}, which lists the cylinder geometry, our estimate of the center-spin magnetization, $m$, the largest MPS bond dimension we reach, $d$, and the associated truncation error, $\epsilon_t$.
For each lattice, we are able to reach good convergence on two or three cylinder sizes.
For the magnetization, we compute the DMRG ground states for an increasing sequence of $d$ and estimate the $d\to\infty$ limit by a linearly extrapolation of the truncation error, $\epsilon_t\to0$.
The linear fit error is displayed when that error is in the third significant digit; smaller errors are physically irrelevant, i.e. compared to finite-size effects.
The LSWT results are within a few percent of the DMRG values. 

\subsection{Spin susceptibility}\label{sec:hydro_susc}
\begin{figure}
    \includegraphics{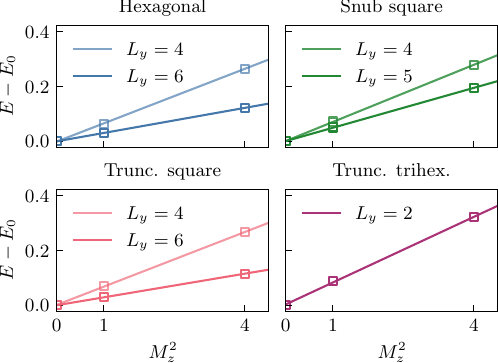}
    \caption{
    Anderson tower of states spectrum on finite cylinders with $U(1)$ conservation.  The slope of each fitted line is $(2 N \chi_\perp)^{-1}$, with the corresponding value of the susceptibility $\chi_\perp$ reported in Table~\ref{tab:susc}.
    }
\end{figure}

\begin{table}
\caption{\label{tab:susc}Spin susceptibility of finite cylinders.}
\begin{ruledtabular}
\begin{tabular}{lll D{.}{.}{2.2} D{.}{.}{1.8} r l}
Lattice & $L_x$ & $L_y$ & \multicolumn{1}{c}{$W$} & \multicolumn{1}{c}{$\chi_\perp J$} & \multicolumn{1}{c}{$d$} & \multicolumn{1}{c}{$\epsilon_t$}\\  
\hline
Hexagonal      & 6 & 4 &  6.93 & 0.1565(12) & 1024 & $4 \times 10^{-8}$ \\
               & 9 & 6 & 10.39 & 0.1512(1)  & 2048 & $1 \times 10^{-6}$ \\ 
                  & & & \multicolumn{1}{c}{LSWT} & 0.1646 & &  \\
\hline
Snub square    & 4 & 4 &  7.73 & 0.1120(10) & 1024 & $6 \times 10^{-6}$ \\
               & 5 & 5 &  9.66 & 0.1026(10) & 2048 & $1 \times 10^{-5}$ \\ 
\hline
Trunc. square  & 4 & 4 &  9.66 & 0.1172(14) & 1024 & $7 \times 10^{-7}$ \\
               & 6 & 6 & 14.49 & 0.1202(1)  & 2048 & $7 \times 10^{-6}$ \\ 
                & & & \multicolumn{1}{c}{LSWT} & 0.162 & &  \\
\hline
Trunc. trihex. & 2 & 2 &  9.46 & 0.1301(30) & 2048 & $6 \times 
10^{-7}$ \\
                  & & & \multicolumn{1}{c}{LSWT} & 0.1588 & &  \\

\end{tabular}
\end{ruledtabular}
\end{table}

The second parameter we determine is the spin susceptiblity, $\chi_\perp$.
Our DMRG approach here is to compute the Anderson tower of states (TOS), that is, the lowest energy state of a finite-size system ($N$ total spins) in each $M^z$ sector~\cite{tasakiLongRangeOrderTower2019}.
The zero-momentum, collective rigid rotor motion of the spins splits the energy of these states by $E_{\mathrm{TOS}} = M_z^2/2 I$, with an extensive moment of inertia $I= \chi_\perp N$~\cite{metlitskiEntanglementEntropySystems2015}.
Thus, we return to using $U(1)$-conserving tensors, compute ground states in the $M_z = 0,1,2$ sectors, and fit their energies to the quadratic relation $E(M_z) = E_0 + M_z^2/(2 N \chi_\perp).$
The results are given in Table~\ref{tab:susc}, with the errors representing the standard error of the fit. 
In this case the LSWT values are 10-30\% larger (more classical) than the DMRG ones.
%




\subsection{Spin stiffness}\label{sec:hydro_stiff}

\begin{figure}\label{fig:stiff}
    \includegraphics{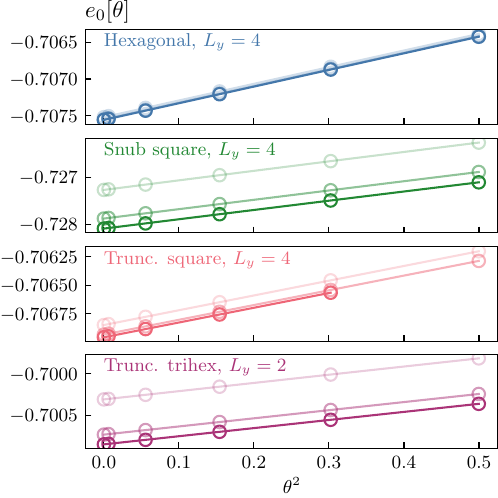}
    \caption{
    Ground state energy density $e_0$ on the infinite cylinder as a function of inserted $U(1)$ flux, $\theta$ (Eq.~\ref{eq:flux}). Different transparencies show energies at increasing MPS bond dimension $d$=512, 1024, 2048. Curves are fits to $e_0 = a_0 + a_2 \theta^2 + a_4\theta^4$, and the corresponding stiffness $\rho = 2 a_2/W^2$ is reported in Table~\ref{tab:stiffness}.
    }
\end{figure}

\begin{table}
\caption{\label{tab:stiffness}Spin stiffness of infinite cylinders.}
\begin{ruledtabular}
\begin{tabular}{lld|d|ll}
Lattice & $L_y$ & W & \rho/J& $d$ & $\epsilon_t$\\
\hline
Hexagonal & 4 & 6.93 & 0.2186(3) & 2048 & $1 \times 10^{-7}$ \\
                  & & \multicolumn{1}{c|}{LSWT} & 0.148 & &  \\
\hline 
Snub square & 4 & 7.73 & 0.236(2) & 2048 & $3 \times 10^{-6}$ \\
\hline 
Trunc. square & 4 & 9.66 & 0.2417(1) & 2048 & $9 \times 10^{-7}$ \\                 
&  & \multicolumn{1}{c|}{LSWT} & 0.142 & &  \\
\hline 
Trunc. trihex. & 2 & 9.46 & 0.1764(4) & 2048 & $5 \times 10^{-6}$ \\
 & & \multicolumn{1}{c|}{LSWT} & 0.134 & &  \\

\end{tabular}
\end{ruledtabular}
\end{table}

The final quantity we measure is the spin stiffness---the energy cost of twisted boundary conditions~\cite{melkoAspectratioDependenceSpin2004}.
For this we return to iDMRG, and carry out a flux insertion experiment~\cite{heSignaturesDiracCones2017}.
Explicitly, we modify couplings of the Hamiltonian according to a $U(1)$ flux $\theta$ through the infinite cylinder,
\begin{equation}\label{eq:flux}
    H_{\rm{dXY}}[\theta] = J\sum_{i<j} \frac{1}{r_{ij}^3} \left(e^{i \theta_{ij}} S_i^+ S_j^- + \mathrm{h.c.}\right)
\end{equation}
where the relative angle $\theta_{ij} = (\mb{r}_{ij}\cdot\mb{W}/W^2)\theta$, with $\mb{W}$ the periodic vector around the cylinder.
Varying $\theta$, the energy dependence  defines the spin stiffness as
\begin{equation}\label{eq:stiffness}
    \rho = W^2 \frac{\partial^2 e_0}{\partial \phi^2},
\end{equation}
where $e_0[\theta]$ is the energy per spin of the $\theta$-dependent ground state.

We compute $e_0[\theta]$ for one width per each cylinder, shown in Fig.~\ref{fig:stiff}.
The corresponding stiffness $\rho$ is listed in Table~\ref{tab:stiffness}.
In this case, the reported errors indicate the difference between calculations done at $d=2048$ and $d=1024$, which is much larger than the uncertainty in the fit to $e_0[\theta]$.
We also include the LSWT prediction, which is notably lower than the DMRG calculation; the largest difference is for the truncated square lattice, at about 40\%. 

\section{Local singlet states}\label{sec:singlet}

\begin{figure}
    \includegraphics{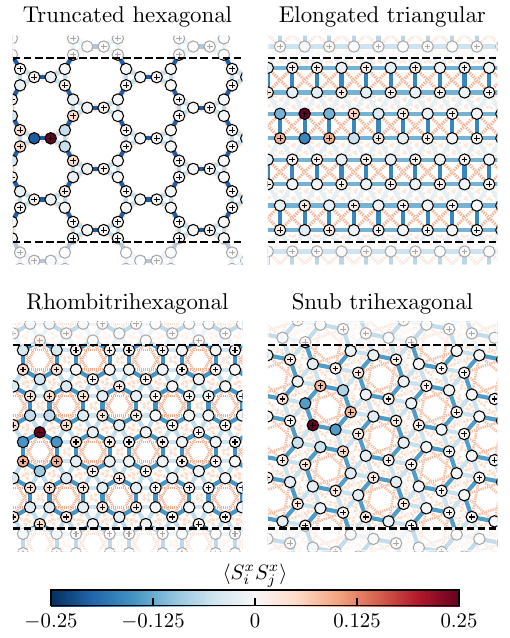}
    \caption{\label{fig:singlets} Depiction of the local-singlet ground states $H_{\rm{dXY}}$ on four of the Archimedean lattices, as found in iDMRG ($d=2048$).  Bonds show correlations between nearest-neighbor and next-nearest neighbors spins; red dashed lines correspond to ferromagnetic $\langle S^x_i S^x_j\rangle>0$, while blue solid lines are antiferromagnetic.  Circles show $\langle S^x_i S^x_0\rangle$ correlations with a fixed site (dark red circle), and those with $\langle S^x_i S^x_0 \rangle >0$ are also marked with a '$+$' sign.     
    }
\end{figure}

Local geometric frustration penalizes collinear ordering.
For some lattices with complex unit cells, we find that the ground state of $H_{\rm dXY}$ is a short-range-entangled paramagnet, where no ordering occurs at all.
These are the truncated hexagonal (``Fisher''), elongated triangular (``isosnub'' or ``trellis''), rhombitrihexagonal (``ruby''), and snub trihexagonal (``maple leaf'') lattices.
On these geometries, neighboring spins within a unit cell become strongly entangled into either two-spin or six-spin singlet states (\textit{a la} benzene).
These local singlets can tile across the whole plane, and the full many-body wavefunction is, to good approximation, a tensor product of them~\cite{sriramshastryExactGroundState1981, ghoshAnotherExactGround2022}.
Let us give three brief remarks on these local singlet states.
First, this type of state is similar to a valence bond solid (VBS), but strictly speaking is not a true ``solid'' as no spatial symmetries are spontaneously broken.
Second, this fully symmetric paramagnetic state is possible because the unit cells contain an even number of spins.
Finally, we note that although our Hamiltonian is only $U(1)$ symmetric, the ground state of an antiferromagnet XY interaction between two spin-1/2 is identical to an $SU(2)$ singlet; this relation no longer holds exactly for our many-body wavefunctions, but we do find that the local correlations in this class of states are largely spin-isotropic (not shown).
We comment on possible directions for further study of these paramagnets in Sec.~\ref{sec:discussion}.

\section{Triangular lattice}\label{sec:tri}

\begin{figure}
    \includegraphics{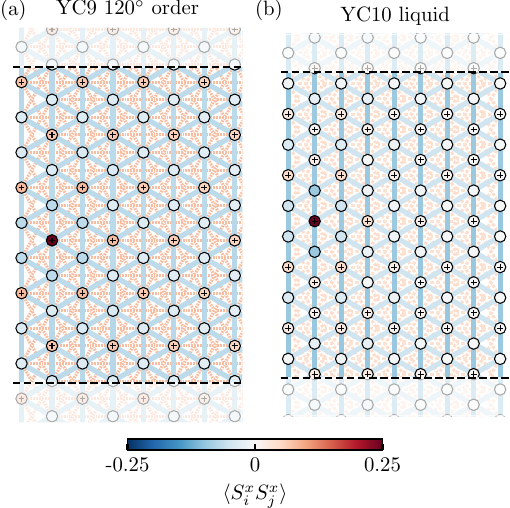}
    \caption{\label{fig:tri_910} 
    Local spin correlations in the triangular lattice iDMRG ground state, following the conventions of Fig.~\ref{fig:collinear}. (a) YC9 cylinder, with long-range 120$^\circ$ magnetic order. (b) YC10 cylinder, with liquid-like short-range correlations.
    }
\end{figure}

We now turn to the triangular lattice.
Collinear order is again disfavored by geometric frustration, but on this geometry there is now no possibility of a trivial local singlet state~\cite{liebTwoSolubleModels1961,oshikawaCommensurabilityExcitationGap2000a,hastingsLiebSchultzMattisHigherDimensions2004}.
One must then consider more intricate states such as non-collinear magnetic orders, valence bond solids, and quantum spin liquids.
This can lead to two complications: (i) enlarged unit cells of the ordering, possibly incompatible with the numerical geometry, (ii) delicate low-energy competition between phases.
Unfortunately, we find in our numerics that each rears its head here, and the second is further complicated by the long-range tail of the dipolar couplings. 
We thus present results showcasing the various competing phases, and leave final judgement between them as a challenge for future investigation.
As a preview, in Fig.~\ref{fig:tri_910} we show the iDMRG spin correlations on the $W=9$ and $W=10$ cylinders; the first has strong symmetry-breaking correlations, while for the latter the correlations quickly decay, suggesting a possible quantum spin liquid.

\subsection{Extended phase diagram}

\begin{figure*}
\includegraphics{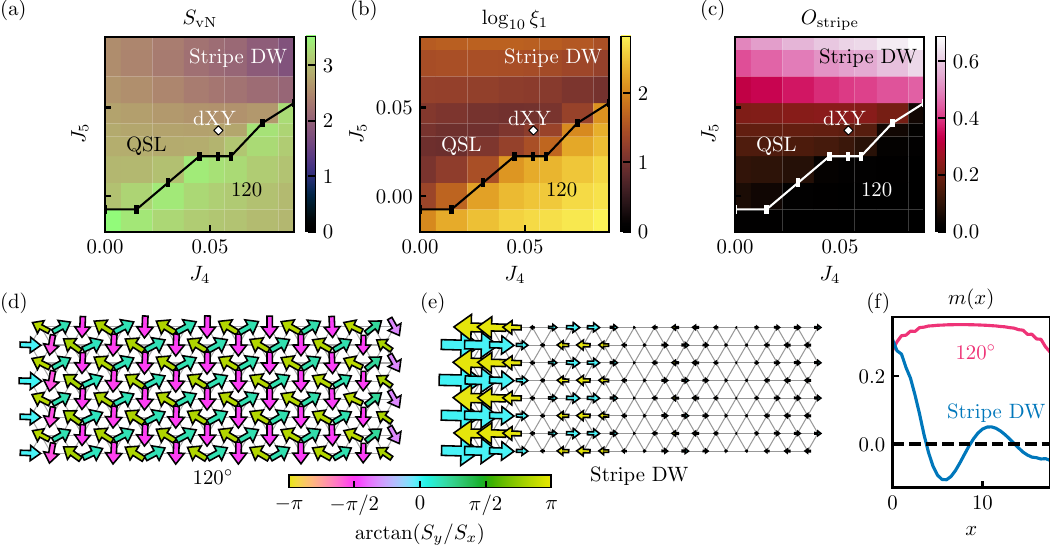}
\caption{\label{fig:global_fiveJtri} Phases of the five-coupling  model on the XC8 triangular lattice cylinder. (a-c) iDMRG phase diagram, fixing $(J_2, J_3) = (0.192, 0.125)$ to match the dipolar Hamiltonian. White diamond marks the dipolar point. Background color measures (a) half-system von Neumman entanglment entropy, $S_\mathrm{vN}$,  (b) spin-1 transfer matrix correlation length, $\xi_1$, (c) spatial anisotropy of correlations, $O_{\mathrm{stripe}} = \langle X_i X_{i+a}\rangle -\langle X_iX_{i+b}\rangle$ where $a=(1,0)$ and $b=(1/2,\sqrt{3}/2)$ are the basis vectors of the triangular lattice.  Black markers show phase boundary inferred from the behavior of $S_{\rm{vN}}$ and $\xi_1$.   (d) Finite DMRG ground state with 120$^{\circ}$ order, at $(J_4, J_5)=(0.054,0)$. We break $U(1)$ and time-reversal symmetry with a pair of weak fields near the edge, $0.2 S_0^x + 0.005S^y_{0+b}$. The resulting local magnetization pattern $(S_i^x, S_i^y)$ is depicted with arrows. (e) Same, but for the stripe density wave (DW) phase, at $(J_4, J_5)=(0.09, 0.09)$.  The arrows in this panel are scaled up by a factor of two, so that some reduced values in the bulk are visible. (f) Local magnetization profile $m_i = \langle \mb{S}_i\cdot\mb{v_i} \rangle$  with respect to in-plane ordering vector $\mb{v}_i = (\cos\theta_i, \sin\theta_i,0)$ for the 120$^{\circ}$ and stripe states in (d,e).  The 120$^{\circ}$ order saturates in the bulk, whereas the stripe order exhibits decaying oscillations with period $\approx 10$ sites and decay length $\approx 6$ sites.
}
\end{figure*}

First, we locate $H_{\rm{dXY}}$ in a broader phase diagram.
Although there are no explicit free parameters in the physical Hamiltonian,  numerically one can always tune the relative strengths of the various couplings.
That is, we consider general XY models of the form,
\begin{equation}
    H_{J} = \sum_{m}  J_m \sum_{i<j} \delta(r_{ij}-d_m) \left[S_i^+ S_j^- + S_i^- S_j^+\right]
\end{equation}
where $d_m$ is the distance corresponding to $m$th-nearest-neighbors on the lattice (e.g. $d_1=1$ for nearest-neighbors, $d_2=\sqrt{3}$ for next-nearest-neighbors, etc.).  
The dipolar model $H_{\rm{dXY}}$ obtains when these couplings decay cubically with distance, i.e.  $J_m = J d_m^{-3}$.
The nearest-neighbor XY AFM is likewise $J_1=J$, $J_{m>1}=0$.

In Fig.~\ref{fig:global_fiveJtri}, we show part of the $H_J$ phase diagram close to the $H_{\rm{dXY}}$ point, as determined by iDMRG. 
We fix the first three couplings to their dipolar values, $(J_1, J_2, J_3) = (1, 0.192, 0.125)$,  and then tune $J_4,J_5$ in the area surrounding the dipolar point $(0.054, 0.037)$.
The longer-range couplings are truncated, $J_{m>6}=0$.
For this parameter scan we use the infinite XC8 cylinder, which is small enough  ($W=4\sqrt{3}\approx6.92$) to get good iDMRG results without prohibitive computational cost ($d=3072$),  while also being commensurate with the unit cells of many common orders.
The half-system von Neumann entanglement entropy [Fig.~\ref{fig:global_fiveJtri}(a)], correlation length [Fig.~\ref{fig:global_fiveJtri}(b)], and spatial anisotropy of local correlations [Fig.~\ref{fig:global_fiveJtri}(c)] suggest the presence of three phases all in close proximity to the rather ambiguously located $H_{\rm{dXY}}$ point.
First, for larger $J_{4}$ and smaller $J_{5}$, we find coplanar 120$^\circ$ magnetic order; the classical order parameter is revealed in a finite DMRG calculation $(d=1024)$ where we add a pair of weak symmetry-breaking local fields to the edge of a cylinder~\cite{whiteNeelOrderSquare2007}, and observe nonzero magnetization, $\langle S_i^x\rangle$ and $\langle S_i^y\rangle$, throughout the bulk [Fig.~\ref{fig:global_fiveJtri}(d,f)].
In iDMRG, we use  $U(1)$-conserving, real-coefficient MPS, for which the corresponding quantum state is a uniform superposition over all $U(1)$ symmetry breaking directions, as well as over the two possible chiralities for the ordering.
This state is readily distinguished in the iDMRG phase diagram by its large correlation length, indicative of long-range off-diagonal order in $S^+S^-$ [Fig.~\ref{fig:global_fiveJtri}(b)].
Second, for smaller $J_{4}$  we find a state with relatively uniform short-range correlations but substantial quantum entanglement.
We do not characterize this state in detail here, but we conjecture it is connected to the well-studied quantum spin liquid (QSL) phase of the triangular lattice $J_1$-$J_2$ Heisenberg model~\cite{zhuSpinLiquidPhase2015, iqbalSpinLiquidNature2016, huDiracSpinLiquid2019, shermanSpectralFunction12023, drescherDynamicalSignaturesSymmetrybroken2023a, jiangCompetingStates$S12026, kovalskaRevisiting$J_1$$J_2$Heisenberg2026}.

Finally, when $J_4$ and $J_5$ are increased, there is a smooth crossover to a lower-entanglement state with spatially-anisotropic spin correlations, suggestive of the collinear stripe order seen in the $J_1$-$J_2$ Heisenberg model.
Curiously, the stripe correlations in our extended XY model are relatively short-ranged, and for the finite DMRG calculation we find that the ordering direction oscillates with a period of about ten sites [Fig~\ref{fig:global_fiveJtri}(e,f)].
We note this stripe density wave (DW) is reminiscent of the mescoscopic ordering patterns seen in thin films with magnetic dipole-dipole interactions~\cite{debellDipolarEffectsMagnetic2000}.

\subsection{Luttinger-Tisza analysis}\label{sec:tri_LT}

To gain some analytic insight into this phase diagram, we use the Luttinger-Tisza method to find the classical ground state of $H_J$~\cite{supp}. 
Since the triangular lattice is a Bravais lattice, the interaction matrix $\tilde{J}(\bm{k})$ simplifies to a single real-valued function,
\begin{equation}
    \tilde{J}(\bm{k};J_4,J_5) = \sum_{n=1}^5 J_n \left[ \sum_{i}\cos({\bm{k}\cdot\bm{\delta}_i^{(n)}})\right],
\end{equation}
where $\bm{\delta}_i^{(n)}$ denote the vectors connecting the $n^\text{th}$ neighbors. 
The ordering wavevector corresponds to the value $\bm{k}_{\mathrm{LT}}$ which minimizes $\tilde{J}(\mb{k})$.
If this momentum is shifted away from a high-symmetry point, the classical ground state takes the form of an in-plane spiral, i.e., $\bm{S}(\bm{R}_i) = \left(\cos{(\bm{k}_{\mathrm{LT}}\cdot\bm{R}_i)}, \sin{(\bm{k}_{\mathrm{LT}}\cdot\bm{R}_i)},0\right)$. 

\begin{figure}
    \centering
    \includegraphics{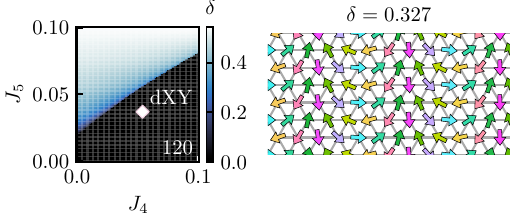}
    \caption{Left: the Luttinger-Tisza phase diagram of the triangular lattice as $J_4 , J_5$ are tuned. The $120^\circ$ AFM order appears in the lower region and transitions into an incommensurate order.
    This has wavevector $\bm{k}_{\mathrm{LT}} = (1-\delta)\bm{K} + \delta  \bm{M}$ that interpolates between $\bm{K}$ and $\bm{M}$ by fraction $\delta$ (colormap).
    Right: Depiction of the spin texture for a representative point of the incommensurate state ($\delta=0.327$).}
    \label{fig:tri_LT}
\end{figure}

In Fig.~\ref{fig:tri_LT}, we show this classical phase diagram surrounding the dipolar point, fixing the first three couplings to their dipolar values, $(J_1, J_2, J_3) = (1, 0.192, 0.125)$, and then tuning $J_4, J_5$ near $(0.054, 0.037)$.
We find that $\bm{k}_{\mathrm{LT}}$ lies on a line between the $K$ and $M$ points of the Brillouin zone.
The $K$-ordered phase in the lower-right half of the phase diagram, encompassing the dipolar point, is the familiar $120^\circ$ AFM.
The upper-left half of the diagram hosts an incommensurate stripe phase. 
We think it would be interesting for future work to clarify the relation between this classical order and the quantum state found in DMRG.

\subsection{Wider cylinders}

\begin{table}
\caption{\label{tab:tri}Triangular lattice $H_{\rm{dXY}}$  ground states (iDMRG).}
\begin{ruledtabular}
\begin{tabular}{ldl|ll|ll}
Cyl. & W & $J_n$ & $120^\circ$ comm. & Order & d & $\epsilon_t$\\
\hline
XC8 & 6.92  & 5 & Yes & Liquid? & 3072 & $4\times10^{-6}$ \\
YC8 &  8 & 7 & No & Liquid & 6144 & $6\times 10^{-6}$ \\
YC9 &  9 & 9 & Yes & $120^\circ$ & 2048 & * \\
YC10 &  10 & 10 & No & Liquid & 6144 & $2\times10^{-5}$ \\
YC8-1  &  7.55 & 6 &  Yes & $120^\circ$  & 2048 & * \\
YC10-1  &  9.54 & 10 & No & Liquid & 3072 & * \\
\end{tabular}
\end{ruledtabular}
\end{table}

\begin{figure}
    \includegraphics{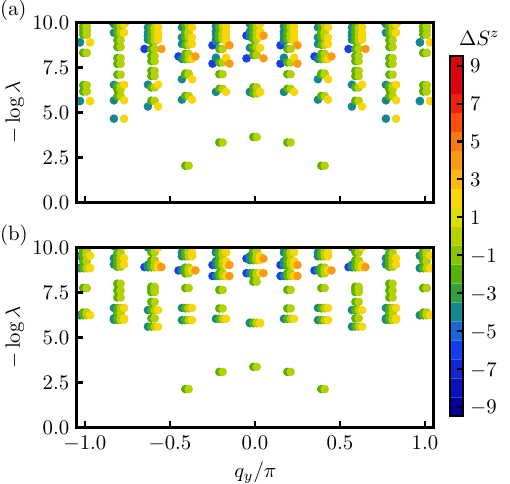}
    \caption{\label{fig:tri_es} 
   Entanglement spectrum of the YC10 liquid state.  Markers correspond to (logarithms of) eigenvalues of the half-system reduced density matrix, sorted by symmetry quantum numbers: the momentum around the cylinder $q_y$, and the spin transfer $\Delta S^z$ (colorbar).  For visual clarity, we slightly offset markers horizontally according to their $\Delta S^z$.  (a) Spectrum for $H_{\rm{dXY}}$. (b) Spectrum for the $J_1$-$J_2$ Heisenberg model at $J_2/J_1=0.12$.  Both exhibit a distinctive arc feature below a denser continuum.
    }
\end{figure}

%
Given the murky situation suggested by the XC8 calculations, it is natural to test whether moving towards the thermodynamic limit ($W\to\infty$, all couplings included) affords any clarity.
To this end, we calculate the iDMRG ground state on several wider cylinders~\cite{szaszChiralSpinLiquid2020}.
In this case tuning across a phase diagram is substantially more expensive, so we restrict our focus to just the state that occurs at the dipolar point (while truncating $J_m=0$ when $d_m\ge W/2$).
Our results are summarized in Table~\ref{tab:tri}, which lists the cylinder geometry, its width, $W$, the number of included couplings $J_n$; whether the 120$^{\circ}$ order is commensurate, the observed ground state order; the MPS bond dimension, $d$, and truncation error, $\epsilon_t$.
(For some cylinders, marked with an asterisk, we encountered better numerical stability using the single-site DMRG algorithm, which does not directly output a truncation error.)
%
%
%
In general, we find that when the 120$^{\circ}$ order is commensurate, it stabilizes; when it is not, a disordered liquid results (Fig.~\ref{fig:tri_910}).
As a first glimpse towards the nature of this liquid, we compute its momentum- and spin-resolved entanglement spectrum~\cite{pollmannDetectionSymmetryprotectedTopological2012, cincioCharacterizingTopologicalOrder2013a} on the YC10 cylinder (Fig.~\ref{fig:tri_es}).
Interestingly, we find an arc-like structure similar to that seen in the $J_1-J_2$ Heisenberg spin liquid~\cite{saadatmandSymmetryFractionalizationTopological2016}.
This is not definitive evidence of topological character \textit{per se}, but it does indicate that the local entanglement properties of the two states are similar, and suggests that they may be in the same phase.  
We leave a more detailed characterization to future work.

\section{Outlook}\label{sec:discussion}
In this work, we characterized the ground states of a dipolar XY Hamiltonian on nine of the eleven Archimedean lattices; the classification is completed by the square and kagome lattices in other work~\cite{chenContinuousSymmetryBreaking2023, sbierskiMagnetismTwodimensionalDipolar2024, bintzDiracSpinLiquid2024}.
Broadly, we found that the lattices with lower geometric frustration (e.g., no frustration at the nearest neighbor level) are typically not significantly affected by the longer range of the dipolar interactions.
Consequently, these lattices stabilize the conventional bipartite N\'eel antiferromagnet as the ground state. 
Furthermore, linear spin-wave theory is able for these to provide a good match to the numerics.
Looking forward, we note that for these ordered phases one can go beyond mean-field linear spin-wave theory and, for example, analytically extract the spin-wave decay-rates~\cite{zhitomirskyColloquiumSpontaneousMagnon2013,zhitomirskyInstabilityAntiferromagneticMagnons1999,kimFieldInducedMagnonDecays2025}. 
This would potentially be directly relevant for experiments, which can locally probe the real-time dynamics of spin-waves, e.g. following a quantum quench~\cite{chenSpectroscopyElementaryExcitations2025a}.

There are also three lattice geometries that we believe are deserving of further study; all three are frustrated at the nearest-neighbor level.
First, the snub square lattice, which we found to be a N\'eel state in DMRG, but possibly an ultimately unstable one as indicated by our analytic calculations.
Second, the snub trihexagonal (maple leaf) lattice, which in our computation seems to have a trivial paramagnet ground state, but in other works (with different Hamiltonians), is considered to be a likely host for quantum spin liquids~\cite{sonnenscheinCandidateQuantumSpin2024, ebertCompetingParamagneticPhases2026}.
Finally, the ordinary triangular lattice: on this geometry, the dipolar Hamiltonian lies in a region of tight competition between three complex phases. 
For each of these geometries, we think it would be interesting to apply different numerical methods, and develop more sophisticated analytic theories.
To conclude, we note from the experimental side there are various accessible perturbations that could induce novel physics on any of the eleven Archimedean lattices.
One pragmatic direction is to investigate whether new phases appear in the presence of a uniform magnetic field~\cite{keselman$J_1J_2$TriangularLattice2025}.
A more distinctive possibility of ultracold atom and molecule experiments, compared to the solid state, is the ability to strongly drive the system with time-varying and site-specific fields~\cite{bukovUniversalHighfrequencyBehavior2015, geierFloquetHamiltonianEngineering2021, bornetEnhancingManybodyDipolar2024a, tianEngineeringFrustratedRydberg2025, luProbingCoherentManyBody2026}.
Could one use this to stabilize non-equilibrium steady states with novel emergent properties~\cite{sahayQuantumSpinPuddles2022}, even when the equilibrium ground states is a conventional antiferromagnet or a featureless paramagnet?
Last but not least, because the lattice geometry itself is highly tunable in these experiments, it seems quite natural to study the effects of controlled geometric deformations on the ground state of the unadorned dipolar XY Hamiltonian~\cite{starykhUnusualOrderedPhases2015a,naygaMagnonLandauLevels2019, youControlDipolarDynamics2025}.

\emph{Acknowledgments.}---We are grateful for illuminating conversations with G. Bornet, A. Browaeys, C. Chen, L. Christakis, G. Emperauger, S. Kivelson, A. D. Kim and T. Lahaye.
This work was supported in part by the U.S. Department of Energy, Office of Science, National Quantum Information Science Research Centers, Quantum Systems Accelerator and by the Air Force Office of Scientific Research via the MURI program (FA9550-21-1-0069).
V.L.~acknowledges support from the NSF through the Center for Ultracold Atoms.
M.Z.~was  supported primarily by the U.S. Department of Energy, Office of Science, Basic Energy Sciences, under Early Career Award No. DE-SC0022716.
N.Y.Y.~acknowledges support from a Simons Investigator award. 
\bibliography{DipolarXY, footnotes_and_other}

\clearpage
\newpage

\onecolumngrid

\appendix






\section{Details on linear spin-wave theory}
In this section, we provide the derivation of the spin-wave Hamiltonian in the canted reference frame when an out-of-plane magnetic field $h$ is present.
To this end, we first re-write the Hamiltonian $H = H_{\rm dXY} - h \sum_i S_i^{z_0}$, given in the lab-fixed frame $(x_0,y_0,z_0)$, in the twisted reference frame which is rotated by an angle $\theta_\eta = \sigma_\eta \theta$ about the $y_0$-axis ($\sigma_\eta = 1$ for type A, and $\sigma_\eta = -1$ for type B sublattice points.)
We fix the angle $\theta$ by demanding that the effective field along the (rotated) $z$ direction vanishes in the Hamiltonian re-written in the twisted reference frame:
\begin{align}
H =&  J \sum^\prime_{(n,\eta),(n^\prime,\eta^\prime)} \frac{S^x_{n,\eta}S^x_{n^\prime,\eta^\prime} \cos \theta_\eta \cos \theta_{\eta^\prime} + S^z_{n,\eta}S^z_{n^\prime,\eta^\prime} \sin \theta_\eta \sin \theta_{\eta^\prime} + S^y_{n,\eta}S^y_{n^\prime,\eta^\prime} - 2 S^z_{n,\eta}S^x_{n^\prime,\eta^\prime}  \sin \theta_\eta \cos \theta_{\eta^\prime} }{|\R_n + \u_\eta - \R_{n^\prime} - \u_{\eta^\prime}|^3} \nn
 & ~~~ - h \sum_{(n,\eta)} (S^z_{n,\eta} \cos\theta_\eta  + S^x_{n,\eta} \sin\theta_\eta)
\label{eq:H2}
\end{align}   
Collecting the terms proportional to $S^z_{n,\eta}$ and taking the classical expectation value of the remaining operators involved (i.e., $S^x_{n,\eta^\prime} \to \langle S^x_{n,\eta^\prime} \rangle =  \sigma_{\eta^\prime} S$ and $ S^z_{n,\eta^\prime} \to \langle S^z_{n,\eta^\prime} \rangle = 0$), we can find the effective field along the $z$ direction in the twisted reference frame.
\begin{align}
H \supset \sum_{n,\eta} h^{\rm eff}_{n,\eta}  S^z_{n,\eta}, \text{ where }
  h^{\rm eff}_{n,\eta} =  - 2 J \sum^\prime_{n^\prime,\eta^\prime}  \frac{\langle S^x_{n^\prime,\eta^\prime} \rangle \sin\theta_\eta \cos\theta_{\eta^\prime}}{|\R_n + \u_\eta - \R_{n^\prime} - \u_{\eta^\prime}|^3} - h \cos \theta_\eta  = \left( 2 J S \ve_0 - h \right) \cos\theta
\end{align}
where we have used that $\langle S^x_{n^\prime,\eta^\prime} \rangle =  \sigma_{\eta^\prime} S$, $\cos(\theta_\eta) = \cos(\theta)$ and $\sin(\theta_{\eta}) = \sin(\sigma_\eta  \theta) =  \sigma_\eta \sin(\theta)$, and also used the definition of the dipolar sum $\ve_0$ from the main text.
This gives us the expression for the canting angle $\theta$ via setting $h^{\rm eff}_{n,\eta} = 0$, yielding 
\begin{equation}
\sin\theta =\frac{h}{2 J S \ve_0},
\end{equation}
which is Eq.~\eqref{eq:sth} in the main text. 
On choosing the canting angle $\theta$ as above, the Hamiltonian in the twisted reference frame takes the following form
\begin{equation}
H = J \sum^\prime_{(n,\eta),(n^\prime,\eta^\prime)}  \frac{S^x_{n,\eta}S^x_{n^\prime,\eta^\prime} \cos \theta_\eta \cos \theta_{\eta^\prime} + S^z_{n,\eta}S^z_{n^\prime,\eta^\prime} \sin \theta_\eta \sin \theta_{\eta^\prime} + S^y_{n,\eta}S^y_{n^\prime,\eta^\prime}  }{|\R_n + \u_\eta - \R_{n^\prime} - \u_{\eta^\prime}|^3} - h \sum_{(n,\eta)} S^x_{n,\eta} \sin\theta_\eta
\label{eq:H3}
\end{equation}  
Now, we express the Hamiltonian in Eq.~\eqref{eq:H3} in terms of linearized Holstein-Primakoff bosons defined as follows: 
\begin{equation}
S^x_{n,\eta} = \sigma_\eta(S - a^\dagger_{n,\eta} a_{n,\eta}), ~ S^y_{n,\eta} = \frac{\sqrt{2S}}{2i}(a^\dagger_{n,\eta} - a_{n,\eta})\sigma_\eta, ~ S^z_{n,\eta} =  \frac{\sqrt{2S}}{2}(a^\dagger_{n,\eta} + a_{n,\eta})
\label{eq:HPdef2}
\end{equation}
Plugging Eq.~\eqref{eq:HPdef2} into Eq.~\eqref{eq:H3} and neglecting all cubic and higher order terms, we have
\begin{align}
H =& \, J S^2 \cos^2\theta \left( \sum^\prime_{(n,\eta),(n^\prime,\eta^\prime)} \frac{\sigma_\eta \sigma_{\eta^\prime}}{|\R_n - \R_{n^\prime} + \u_\eta - \u_{\eta^\prime}|^3} \right) - J S \cos^2\theta \left( \sum^\prime_{(n,\eta),(n^\prime,\eta^\prime)} \frac{\sigma_\eta \sigma_{\eta^\prime}(a^\dagger_{n,\eta} a_{n,\eta} + a^\dagger_{n^\prime,\eta^\prime} a_{n^\prime,\eta^\prime})}{|\R_n - \R_{n^\prime} + \u_\eta - \u_{\eta^\prime}|^3}\right) \nn 
& - \frac{JS \cos^2\theta}{2} \left(  \sum^\prime_{(n,\eta),(n^\prime,\eta^\prime)}  \frac{\sigma_\eta \sigma_{\eta^\prime}(a^\dagger_{n,\eta} a^\dagger_{n^\prime,\eta^\prime} + a_{n,\eta} a_{n^\prime,\eta^\prime})}{|\R_n - \R_{n^\prime} + \u_\eta - \u_{\eta^\prime}|^3}\right) + \frac{JS (1 + \sin^2\theta)}{2}\left( \sum^\prime_{(n,\eta),(n^\prime,\eta^\prime)}  \frac{\sigma_\eta \sigma_{\eta^\prime}(a_{n,\eta} a^\dagger_{n^\prime,\eta^\prime} + a^\dagger_{n,\eta} a_{n^\prime,\eta^\prime})}{|\R_n - \R_{n^\prime} + \u_\eta - \u_{\eta^\prime}|^3}\right) \nn
& - h \sin\theta \left( \sum_{n,\eta} (S - a^\dagger_{n,\eta} a_{n,\eta}) \right)
\label{eq:HHP}
\end{align}
Note that we do not have any linear term in the HP bosons in the Hamiltonian in Eq.~\eqref{eq:HHP}, the condition that the effective field on any spin vanishes is essentially an energy minimization condition which gets rid of these linear terms. 

Finally, we write the Hamiltonian in Eq.~\eqref{eq:HHP} in terms of the Fourier transformed HP bosons, leading to 
\begin{align}
H &= H^{(0)}_{\rm cAFM} + H^{(2)}, \text{ where} \nn
H^{(0)}_{\rm cAFM} &= -2sN \left[ J S(S + 1) \ve_0 \cos^2\theta + h \left(S + \frac{1}{2}\right) \sin\theta \right] = -2sN \left[ J S(S + 1) \ve_0 + \frac{h^2}{4 J \ve_0} \right] \nn
H^{(2)} & = \frac{J S}{2} \sum_{\k,\eta, \eta^\prime} \left[ 2 \ve_0 \, \delta_{\eta \eta^\prime} - (1 + \sin^2\theta) \ve_{\k,\eta, \eta^\prime} \right] \left(  a^\dagger_{\k,\eta} a_{\k,\eta^\prime}  +   a_{-\k,\eta} a^\dagger_{-\k,\eta^\prime} \right) + \cos^2\theta \, \ve_{\k, \eta \eta^\prime} 
 \left( a_{-\k,\eta} a_{\k, \eta^\prime} +  a^\dagger_{\k, \eta} a^\dagger_{-\k,\eta^\prime}  \right),
\end{align}
which is Eq.~\eqref{eq:HHPquadratic} in the main text.

\begin{figure}
    \centering
    \includegraphics[width=1.0\linewidth]{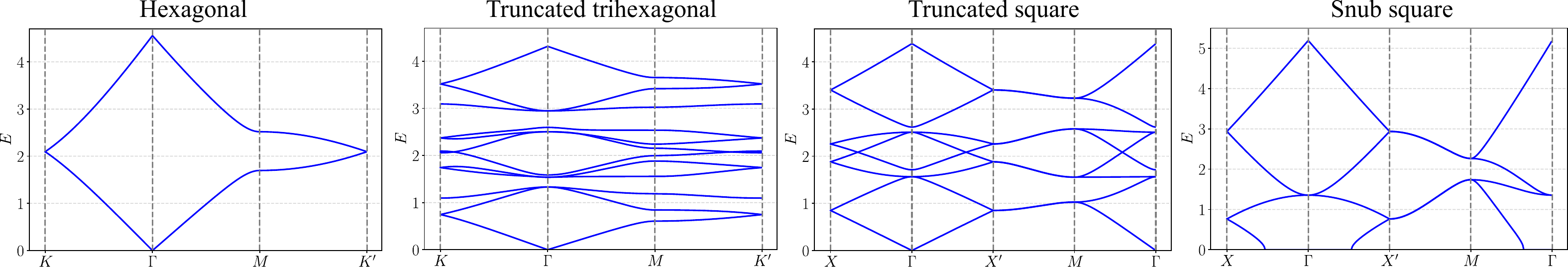}
    \caption{Line-cuts of the spin-wave spectrum of the four lattices where iDMRG finds a ground state with antiferromagnetic two-sublattice N\'eel order. For three of the four lattice, the spin-wave spectrum is well-defined. However, for the snub square lattice (right-most panel), the spin-wave energy drops to zero and becomes ill-defined at low momentum, indicating an instability.}
    \label{fig:PRB-SM-SW_spec}
\end{figure}

Finally, we present the spin-wave spectra $\omega_{\k,\mu}$ for the $2s$ spin-wave branches in Fig.~\ref{fig:PRB-SM-SW_spec}, obtained by diagonalizing $H^{(2)}$ via a paraunitary matrix $W_\k$ as detailed in the main text. 
Evidently, the hexagonal, truncated trihexagonal and truncated square lattices all show an acoustic mode $\mu_0$ that disperses linearly at low-energy, i.e., $\omega_{\k,\mu_0} \approx v_{\mu_0}|\k|$ near the $\Gamma$-point.
However, the snub square lattice shows that the spin-wave energy becomes imaginary near the $\Gamma$-point, indicative of an instability of the N\'eel state as discussed in the main text. 

\section{Out-of-plane susceptibility \texorpdfstring{$\chi_\perp$}{chiperp} in linear spin-wave theory}
Here, we derive the equation for susceptibility of the ordered antiferromagnetic N\'eel phases to an out-of-plane magnetic field $h$.
We note from the main text that the total magnetization $M_{z_0}$ in the $z_0$-direction is given by
\begin{equation}
M_{z_0} = \sum_{(n,\eta)}  \langle  S^{z_0}_{n,\eta} \rangle = \sum_{(n,\eta)} \left( \langle S^z_{n,\eta} \rangle\cos\theta_\eta  + \langle S^x_{n,\eta} \rangle \sin\theta_\eta \right) = \sin\theta \sum_{(n,\eta)} (S - \langle a^\dagger_{n,\eta} a_{n,\eta}\rangle ) = \frac{h}{2 J S \ve_0} \sum_{(n,\eta)} (S - \langle a^\dagger_{n,\eta} a_{n,\eta}\rangle ) 
\label{eq:chiU_SM}
\end{equation}
where we have used that $\langle S^z_{n,\eta} \rangle = 0$ within linear-spin wave theory as it is made of operators that are linear in $a$-bosons. 
The first term in Eq.~\eqref{eq:chiU_SM} is just the classical value, and the second term represents the reduction due to quantum fluctuations. 
Since we want the differential susceptibility as $h \to 0$, i.e., $\chi_\perp \propto \partial_h M_{z_0}|_{h \to 0} $, we may evaluate the average HP boson occupancy $ \langle a^\dagger_{n,\eta} a_{n,\eta}\rangle$ for $h = 0$ since the overall proportionality constant  already scales as $h$.  

To derive a compact expression for the boson occupancy, it is convenient to move to Fourier space. 
Using the transformation $W_\k$ used to diagonalize $H^{(2)}$, we may write $a_{\k,\eta}$ in terms of $\gamma_{\k,\mu}$ and calculate the susceptibility as $h^{(2)}$ is diagonal in the number operator for $\gamma_{\k,\mu}$ bosons, as follows: 
\begin{align}
\sum_{m,\eta} \langle a^\dagger_{m,\eta} a_{m,\eta}\rangle &=   \sum_{\k,\eta} \langle a^\dagger_{\k,\eta} a_{\k,\eta} \rangle = \frac{1}{2}\sum_{\k,\eta} ( \langle a^\dagger_{\k,\eta} a_{\k,\eta} \rangle + \langle a^\dagger_{-\k,\eta} a_{-\k,\eta} \rangle ) \nn &= \frac{1}{2}\sum_{\k,\eta} ( \langle a^\dagger_{\k,\eta} a_{\k,\eta} \rangle + \langle a_{-\k,\eta} a^\dagger_{-\k,\eta} \rangle - 1) \nn
& = \left( \frac{1}{2} \sum_\k \langle \Psi^\dagger_\k \Psi_\k \rangle \right) - \frac{2Ns}{2} \nn
& = \left( \frac{1}{2} \sum_\k \langle \Gamma^\dagger_\k W_\k^\dagger W_\k \Gamma_\k \rangle \right) - Ns \nn
& = \frac{1}{2} \sum_{\k} \left[ \sum_{\nu = 1}^{2s} [W_\k^\dagger W_\k]_{\nu,\nu} ~ n_B(\omega_{\k,\nu}) + \sum_{\nu = 2s+1}^{4s} [W_\k^\dagger W_\k]_{\nu,\nu} \left(1 + n_B(\omega_{\k,\nu}) \right) \right] - Ns
\end{align}
where $[W_\k^\dagger W_\k]_{\nu,\nu}$ indicates the $\nu^{th}$ diagonal element of the $2s \times 2s$ matrix $W^\dagger_\k W_\k$, and we have defined $\omega_{\k,2s+1} \equiv \omega_{\k,s}$.
According to the above expression, out-of-plane magnetization per site is simply $m_{z_0} = M_{z_0}/(2sN)$, and thus the out-of-plane susecptibility to a uniform field from Eq.~\eqref{eq:chiU_SM} per site is given by
\begin{equation}
\chi_\perp = \frac{\partial m_{z_0}}{\partial h}\bigg|_{h \to 0} = \frac{1}{2 J S \ve_0}\left(S + \frac{1}{2} - \frac{1}{4sN} \sum_{\k} \left[ \sum_{\nu = 1}^{2s} [W_\k^\dagger W_\k]_{\nu,\nu} ~ n_B(\omega_{\k,\nu}) + \sum_{\nu = 2s+1}^{4s} [W_\k^\dagger W_\k]_{\nu,\nu} \left(1 + n_B(\omega_{\k,\nu}) \right) \right]  \right)
\label{eq:chiperpT}
\end{equation} 
Finlly, considering the $T = 0$ limit of Eq.~\eqref{eq:chiperpT}, when $n_B(\omega_{\k,\nu}) = 0$ and we simply need the $2s$ lower diagonal elements of the positive semi-definite matrix $W_\k^\dagger W_\k$, we recover Eq.~\eqref{eq:chip} in the main text.

\section{Luttinger-Tisza analysis}
In this section, we provide a brief overview of the Luttinger-Tizsa analysis~\cite{luttingerTheoryDipoleInteraction1946} that is used in the main text.
We begin by representing the quantum XY-model spins as classical two-dimensional vectors with unit length, with the Hamiltoninan taking the general form,
\begin{equation}
    H = \frac{1}{2}\sum_{n,n^\prime}\sum_{\eta,\eta^\prime}J_{n n^\prime}^{\eta,\eta^\prime}\bm{S}_{n,\eta} \cdot\bm{S}_{n^\prime,\eta^\prime},
\end{equation}
where $n,n^\prime$ denote the Bravias lattice positions and $\eta, \eta^\prime$ denote the sublattice labels.
The method proceeds by writing the Hamiltonian in momentum space by defining
\begin{equation}
    \bm{S}_{n,\eta} = \frac{1}{\sqrt{N}}\sum_{\bm{k}}\tilde{\bm{S}}_{\bm{k}}^\eta e^{i\bm{k}\cdot\bm{R}_n}
\end{equation}
The Hamiltonian can thus be written as  
\begin{equation}
    \label{eq:LTHam}
    H = \sum_{\bm{k}}\sum_{\eta,\eta^\prime}\tilde{J}^{\eta \eta^\prime}(\bm{k})\tilde{\bm{S}}_{-\bm{k}}^\eta \tilde{\bm{S}}_{\bm{k}}^{\eta^\prime},
\text{ where }
    \tilde{J}^{\eta \eta^\prime}(\bm{k}) = \frac{1}{2}\sum_{\bm{R}_n} J^{\eta \eta^\prime}(\bm{R}_n) \, e^{i \bm{k} \cdot \bm{R}_n}
\end{equation}
is the Fourier-transformed interaction matrix in the sublattice space. 
This becomes a quadratic form for which one needs to find the spin configuration that minimizes the energy, subject to the unit length constraint on each spin (strong constraint) that makes the problem analytically intractable on most lattices. 
The Luttinger-Tisza insight is to relax this constraint to the weak constraint as a necessary (but not sufficient) condition that the strong constraint is followed, 
\begin{equation}
    \sum_{n,\eta} |\bm{S}_{n\eta}|^2 = N N_s,
\end{equation}
or, equivalently in momentum space:
\begin{equation}
    \sum_{\bm{k}} \sum_\eta |\bm{S}^\eta_{\bm{k}}|^2 = N N_s,
\end{equation}
where $N$ is the number of unit cells and $N_s$ is the number of sublattices in each unit cell. 
The problem of minimization simplifies into an eigenvalue problem, and then one needs to check if the strong constraint is satisfied to declare a solution is found. 
Writing Eq.~\eqref{eq:LTHam} in matrix notation, we have, 
\begin{equation}
    H = \sum_{\bm{k}}\tilde{S}_{\bm{k}}^\dagger\tilde{J}(\k) \tilde{S}_{\bm{k}} = \sum_{\bm{k}}\sum_{\nu=1}^{N_s}\lambda_\nu(\bm{k}) |\bm{\zeta}_{\nu\bm{k}}|^2 
\end{equation}
The energy is minimized while simultaneously satisfying the weak constraint only if the mode with the minimum eigenvalue $\lambda_{\mathrm{min}}(\bm{k}_{LT})$ is allowed. 

We checked numerically that this method gives the N\'eel states as the classical ground states (i.e. the strong constraint is also satisfied) for the the three lattices where LSWT works. 
On the other hand, the sunb square lattice has $\bm{k}_{LT}$ which deviates from the Gamma point and lies at an incommensurate point along the $\Gamma-M$ line (and the other momenta related by the $C_4$ symmetry). 
The minimum wave-vector moving away from the $\Gamma$ point hints at a different classical ground state on the snub square lattice.  
These results are presented in Fig.~\ref{fig:LT_spectrum} in the main text.

\end{document}